\documentclass{iopart}

\usepackage{graphicx}
\usepackage{dcolumn}
\usepackage{bm}
\usepackage{epsf}
\usepackage{amssymb}

\newcommand{\bra}[1]{\langle #1|}
\newcommand{\ket}[1]{|#1\rangle}
\newcommand{\braket}[2]{\left\langle #1|#2\right\rangle}
\newcommand{\spur}[1]{\tr\left\{#1\right\}}

\newcommand{\la}{\langle}
\newcommand{\ra}{\rangle}

\newcommand{\td}{{d}}
\newcommand{\ma}[1]{\max{\left\{#1\right\}}}

\newcommand{\com}[2]{\left[#1,\,#2\right]}

\newcommand{\co}[1]{\cos{\left(#1\right)}}
\newcommand{\si}[1]{\sin{\left(#1\right)}}

\newcommand{\bla}{bla\\bla\\bla\\bla\\bla}
\newcommand{\PRA}{\textit{Phys. Rev. A }}

\newcommand{\PRD}{\textit{Phys. Rev. D }}
\newcommand{\PRE}{\textit{Phys. Rev. E }}

\begin{document}

\title{Energy-time uncertainty relation for driven quantum systems}
\author{Sebastian Deffner$^{1,2}$ and Eric Lutz$^{1,3,4}$}
\address{$^1$Department of Physics, University of Augsburg, D-86135 Augsburg, Germany\\$^2$Department of Chemistry and Biochemistry and Institute for Physical Science and Technology, University of Maryland, 
College Park, Maryland 20742, USA\\$^3$Dahlem Center for Complex Quantum Systems, FU Berlin, D-14195 Berlin, Germany\\$^4$ Institute for Theoretical Physics, University of Erlangen-N\"urnberg, D-91058 Erlangen, Germany}
\ead{sebastian.deffner@gmail.com}
\date{\today}

\begin{abstract}
 We derive  generalizations of the energy-time uncertainty relation for driven  quantum systems. Using a geometric approach based on the Bures length between  mixed quantum states, we obtain explicit expressions for the quantum speed limit time, valid for  arbitrary initial and final quantum states and arbitrary unitary driving protocols. Our results establish the fundamental  limit on the rate of evolution of closed quantum systems.
\end{abstract}

\pacs{03.65.-w, 03.67.Hk}

\maketitle

The Heisenberg uncertainty relations are among the most interesting peculiarities of quantum mechanics. They  express the intrinsic duality between particle and wave aspects of quantum-mechanical objects \cite{aul09}. Unlike for  the uncertainty relation for position and momentum, $\Delta x\, \Delta p \geq\hbar/2$ \cite{hei27,boh28,rob29}, the proper foundation and interpretation  of the  relation for energy and time, $\Delta E\, \Delta t\geq \hbar/2$, has been  a delicate issue (see  Ref.~\cite{bus07} for a  review). The main problem can be traced back to the fact that generally time is only a   parameter in quantum theory and that there is not one well-defined 'time'-operator. For example, Aharonov and Bohm have  shown that, contrary to common belief, $\Delta E$ cannot be regarded as the  minimum dispersion of an energy measurement of duration $\Delta t$ \cite{aha61}. Rather, $\Delta t$ corresponds to the minimum time interval that quantum systems with constant energy and initial energy spread $\Delta E$ need to evolve between two orthogonal states \cite{ana90,vai91,uff93,bro03}. The quantity $\tau_\mathrm{QSL}\simeq \hbar/\Delta E$ thus defines the so-called quantum speed limit time of a system. The first rigorous derivation of the quantum speed limit time goes back to  Mandelstam and Tamm \cite{man45} (see also Refs.~\cite{fle73,bha83}). They have established that, for an undriven quantum system,   $\tau_\mathrm{MT}=\pi\hbar/2 \Delta E $, where $\Delta E=(\la H\ra^2-\la H^2\ra)^{1/2}$. More recently, Margolus and Levitin have obtained another bound on the quantum evolution time, $\tau_\mathrm{ML}=\pi\hbar/2 E$,  in terms of the initial  mean energy of the system, $E=\la H\ra-E_\mathrm{g}$,  with respect to  the ground state \cite{mar98}.  
The energy-time uncertainty relation for time-independent systems has been further extended by Giovannetti, Lloyd, and Maccone \cite{gio03,gio04}, who determined the quantum speed limit time for general mixed states,  not necessarily orthogonal, as a function of their geometrical angle given by the Bures length \cite{bur69}.

The purpose of the present paper is to generalize the energy-time uncertainty relation to driven closed quantum systems. Specifically, we derive the quantum speed limit time for generic mixed states with arbitrary angle and for arbitrary time-dependent unitary driving.  We obtain  both Mandelstam-Tamm  and  Margolus-Levitin type uncertainty relations. To this end, we extend the geometric approach put forward by Jones and Kok  for time-independent systems and orthogonal states \cite{jon10} (see also Refs.~\cite{ana90}). Early work on time-dependent systems can be found in Refs.~\cite{uhl92,pfe93,pfe95}. The quantum speed limit time gives the fundamental limit on the rate of evolution of quantum systems \cite{pfe93,pfe95,lev09}. As such it plays a crucial role in the determination of the maximum rate of quantum communication \cite{bek81}, quantum computation \cite{llo00}, quantum metrology \cite{zwi10,gio11} and  nonequilibrium quantum entropy production \cite{def10}. It  also sets   the inherent speed limit of quantum optimal control algorithms \cite{can09,mur10}. As we show below, the quantum speed limit strongly depends on the external driving.

Geometric concepts are essential  tools in physics, not only in general relativity, but also in quantum theory \cite{ben06}.  The angle in Hilbert space between two pure states (or wave vectors), $\ket{\psi_0}$ and $\ket{\psi_\tau}$, is given by  \cite{ben06},
\begin{equation}
\label{q01}
\ell\left(\psi_0,\psi_\tau\right) =\arccos{\left( \int \td x\, \sqrt{\mathcal{P}_0\left( x\right) \, \mathcal{P}_\tau\left( x\right) } \right) }\ ,
\end{equation}
with the probability distributions $\mathcal{P}_0(x) = |\psi_0(x)|^2$ and $\mathcal{P}_\tau(x) = |\psi_\tau(x)|^2$. Wootters has shown that the geodesic distance, $\ell\left(\psi_0,\psi_\tau\right) =\ell\left(\mathcal{P}_0,\mathcal{P}_\tau\right)$, is equal to the number of distinguishable distributions  between  $\mathcal{P}_0$ and $\mathcal{P}_\tau$  along a given line in parameter space, for example time \cite{woo81} -- two distributions are here said to be distinguishable if their separation is larger than their widths.  Wootters' statistical distance is the only Riemannian metric,
up to a constant factor, which is invariant under all unitary
transformations. The notion of distinguishability is often quantified with the help of the Fisher information \cite{cov06}. Let us consider the infinitesimal distance $\td\ell$ between two probability distributions   $\mathcal{P}_t$ and $\mathcal{P}_t+\td \mathcal{P}_t$.
Since  the distribution $\mathcal{P}_t$ depends on the single parameter $t$, we obtain from Eq.~(\ref{q01}),
\begin{equation}
\label{q02}
(\td_t\, \ell)^2 =\int \td x \frac{(\td_t\,\mathcal{P}_t(x))^2}{\mathcal{P}_t(x)}= \int \td x\, \mathcal{P}_t(x) (\td_t\ln \mathcal{P}_t(x))^2 \ .
\end{equation}
The expression  on the right-hand side defines the Fisher information $f_t$ \cite{cov06}. It is given by the variance of the logarithmic derivative of the distribution with respect to the parameter $t$ and, hence, quantifies its sensitivity to a change of the latter. The Fisher information for a Gaussian distribution,  $\mathcal{P}_t(x) = N_t \exp[-(x-\mu_t)^2/(2\sigma_t^2)]$ is equal to  the inverse of its width, $f_t= 1/\sigma_t^2$. As a result, we have $\ell = \int dt/\sigma_t$.

The angle between two density operators $\rho_0$ and $\rho_\tau$ is given by the Bures length (or Bures  angle) \cite{bur69,kak48},
\begin{equation}
\label{q03}
\mathcal{L}\left(\rho_0,\rho_\tau\right)=\arccos{\left( \sqrt{F\left(\rho_0,\rho_\tau\right)} \right) }\,,
\end{equation}
where $F\left(\rho_0,\rho_\tau\right)$ is the quantum fidelity \cite{uhl76,joz94,nie00}, a mixed-state generalization of the usual overlap of two wave vectors. The Bures length (\ref{q03}) has been shown to be   a natural generalization of Wootters' statistical distance for arbitrary mixed quantum states \cite{bra94}. 
The quantum fidelity is defined as \cite{uhl76,joz94,nie00},
\begin{equation}
\label{q04}
F\left( \rho_0,\rho_\tau\right)=\left[\spur{\sqrt{\sqrt{\rho_0}\,\rho_\tau\, \sqrt{\rho_0}}} \right]^2\,.
\end{equation}
For pure  states, $\rho_0 =|\psi_0 \ra \la \psi_0  |$ and $\rho_\tau =|\psi_\tau \ra \la \psi_\tau  |$, it reduces to their overlap,
$F(\rho_0,  \rho_\tau) = \spur {\rho_0 \rho_\tau} = |\la \psi_0| \psi_\tau \ra|^2$. In this case,  $\mathcal{L}$ reduces to  $\ell$.
Due to the square roots  in Eq.~(\ref{q04}), the fidelity  is in general complicated to handle. 
In the following, we consider  quantum systems whose Hamiltonians $H_t$ are driven by arbitrary time-dependent parameters during time $\tau$. We shall obtain generalizations of the energy-time uncertainty relation by deriving upper bounds for the dynamical velocity, $\td_t\, \mathcal{L}$, defined as the derivative of the Bures length with respect to the parameter $t$  \cite{ana90}. 

\section{Mandelstam-Tamm bound}

We begin with a derivation of a Mandelstam-Tamm type uncertainty relation. For that purpose, we consider the square of the dynamical velocity $(\td_t\, \mathcal{L})^2$. According to Braunstein and Caves, the Bures metric for two infinitesimally close density operators, $\rho'=\rho+\td\rho$, is \cite{bra94} (see also Ref.~\cite{hub92}),
\begin{equation}
\label{q05}
\td \mathcal{L}^2=\spur{\td\rho\, \mathcal{R}_\rho^{-1}(\td\rho)}\ ,
\end{equation}
where the superoperator $\mathcal{R}^{-1}$ for an arbitrary operator $O$ reads in terms of the eigenvalues $p_i$ of $\rho$, $\rho=\sum_i\,p_i\ket{i}\bra{i}$,
\begin{equation}
\label{q06}
\mathcal{R}_\rho^{-1}(O)=\frac{1}{2}\sum_{j,k}\,\frac{\bra{j}O\ket{k}}{p_j+p_k}\ket{j}\,\bra{k} \ .
\end{equation}
Note that the superoperator $\mathcal{R}_\rho^{-1}$ is here defined as describing the infinitesimal Bures angle $\mathcal{L}$, and hence differs by a factor 4 from the one used in Ref.~\cite{bra94}, where $\mathcal{R}_\rho^{-1}$ is determined by the infinitesimal statistical distance.

At the same time,  we can rewrite the von Neumann equation for the density operator of the system in the form,
\begin{equation}
\label{q07}
i\hbar\,\td_t \rho_t=\com{H_t}{\rho_t}=\com{H_t-\la H_t\ra}{\rho_t}=\com{\Delta H_t}{\rho_t}\,,
\end{equation}
since the expectation value of the energy  $\la H_t\ra$ is a real number that can be included in the commutator. Combining Eqs.~(\ref{q05})-(\ref{q07}), we find,
\begin{eqnarray}
\label{q08}
(\td_t\, \mathcal{L})^2&=\spur{\td_t\rho_t\, \mathcal{R}^{-1}_{\rho_t}(\td_t\rho_t)}=\frac{1}{2\hbar^2}\,\sum_{j,k}\,\frac{\left(p_j-p_k\right)^2}{p_j+p_k}\,\left|\bra{j}\Delta H_t\ket{k}\right|^2 \nonumber\\
&\leq\frac{1}{2\hbar^2}\,\sum_{j,k}\,\left(p_j+p_k \right)\,\left|\bra{j}\Delta H_t\ket{k}\right|^2=\frac{1}{\hbar^2}\la( \Delta H_t)^2\ra\,,
\end{eqnarray}
where the last line follows from a triangle-type inequality \cite{bra96}. Note that Eq.~(\ref{q08}) also implies that the Fisher information $f_t$ is bounded from above by the variance of the energy of the system. The generalized energy-time uncertainty relation is now obtained by first taking the  positive root of   Eq.~(\ref{q08}),
 \begin{equation}
\label{q09}
\td_t \mathcal{L}\leq\left| \td_t\,\mathcal{L}\right|\leq\frac{1}{\hbar}\,|\langle (\Delta H_t)^2\rangle^{1/2}| \ ,
\end{equation}
 and then  performing the integral over both the Bures length and time,
\begin{equation}
\label{q10}
\int_0^{\mathcal{L}\left(\rho_0,\rho_\tau\right)} \td\mathcal{L}\leq\frac{1}{\hbar}\,\int_0^\tau\td t\,|\langle (\Delta H_t)^2\rangle^{1/2}|\,.
\end{equation}
As a result, we obtain the inequality,
\begin{equation}
\label{q11}
\tau\geq\frac{\hbar}{\Delta E_\tau}\,\mathcal{L}\left(\rho_0,\rho_\tau\right)\ ,
\end{equation}
where the time averaged energy variance is given by $\Delta E_\tau=(1/\tau)\,\int_0^\tau\td t\,(\la H_t^2\ra -\la H_t \ra^2)^{1/2}$. Equation  (\ref{q11}) is 
a Mandelstam-Tamm uncertainty relation  valid for arbitrary initial and final mixed quantum states and arbitrary time-dependent  Hamiltonians. It has been obtained earlier by Uhlmann using parallel Hilbert-Schmidt operators \cite{uhl92}.

\section{Margolus-Levitin bound}

We next establish a Margolus-Levitin type bound  for the quantum speed limit.  For  simplicity, and  without loss of generality, we assume that the driven quantum system starts   in a pure state  $\rho_0=\ket{\psi_0}\bra{\psi_0}$. Due to the unitary dynamics the system then remains in a pure state at all times. Otherwise, we would have to handle the tedious evaluation of the derivatives of square roots of operators in the argument of the fidelity function $F(\rho_0,\rho_\tau)$ (\ref{q03}). The case of an initial mixed state can be treated by making use of  the concept of purification, that is, constructing mixed states by taking partial traces over pure states in a sufficiently enlarged Hilbert space \cite{joz94}. We additionally  set the ground state energy to zero, $E_g=0$ \cite{mar98}. 
Following Ref.~\cite{mar98}, we consider the overlap of initial and final states and write,
\begin{eqnarray}
\label{q33}
\left|\braket{\psi_0}{\psi_\tau}\right|&=\left|\bra{\psi_0}U_\tau\ket{\psi_0}\right|=\left|\sum\limits_n \left|\braket{\psi_0}{n}\right|^2\exp{\left(-i J_n\right)}\right|\nonumber\\
&\geq\left|\bra{\psi_0}\co{J_\tau}\ket{\psi_0}\right| \,,
\end{eqnarray}
where $U_t$ denotes the time evolution operator and $\{\ket{n}\}$ is the set of its instantaneous eigenstates, with $(1/\hbar)\,\int_0^t\,\td t' H_{t'}\ket{n}\equiv J_t\ket{n}= J_n\ket{n}$. The last inequality follows from the observation that $|a+i b|= \sqrt{a^2+b^2}\geq |a|$. Moreover, by using the series expansion of the cosine, $\co{x}=\sum_\mu(-1)^\mu \,x^{2\mu}/(2\mu)!$, we consider,
\begin{equation}
\label{q34}
\left|\bra{\psi_0}J_\tau^{2\mu}\ket{\psi_0}\right|\geq \left|\bra{\psi_0}J_\tau\ket{\psi_0}^{2\mu}\right|\,,
\end{equation}
where the inequality follows from Jensen's inequality with $\mu$ being a positive integer. Accordingly, we have
\begin{equation}
\left|\bra{\psi_0}\co{J_\tau}\ket{\psi_0}\right|\geq \left|\co{\bra{\psi_0} J_\tau \ket{\psi_0}}\right|\,,
\end{equation}
and since the $\arccos$  is a monotonically decreasing function in the interval $[0,1]$, we finally have,
\begin{eqnarray}
\label{q35}
  \arccos{\left(\left|\braket{\psi_0}{\psi_\tau}\right|\right)}&\leq\arccos{\left(\left|\co{\frac{1}{\hbar}\int_0^\tau \td t\,\bra{\psi_0}H_t\ket{\psi_0}}\right|\right)}\nonumber\\
&\leq \left|\frac{1}{\hbar}\int_0^\tau \td t\,\bra{\psi_0}H_t\ket{\psi_0}\right|\leq\frac{1}{\hbar}\int_0^\tau \td t\,\left|\bra{\psi_0}H_t\ket{\psi_0}\right|\,.
\end{eqnarray}
As a consequence, the minimal quantum evolution time from initial to final state can be estimated from below by
\begin{equation}
\label{q36}
 \tau\geq\frac{\hbar}{E_\tau}\,\mathcal{L}\left(\psi_0,\psi_\tau\right)\,,
\end{equation}
where $E_\tau=(1/\tau)\, \int_0^\tau\td t\, \left|\bra{\psi_0} H_t \ket{\psi_0}\right|$ is the time-averaged mean energy with respect to the initial state.  The generalization to arbitrary mixed quantum states can be done invoking purification and the mean energy becomes $E_\tau=(1/\tau)\, \int_0^\tau\td t\, \left|\spur{\rho_0 H_t}\right| $, cf. also appendix. The corresponding inequality is an extension of the result of Jones and Kok, obtained for constant Hamiltonians and orthogonal states \cite{jon10}. 

\section{Discussion}

Summing up, we have obtained the quantum speed limit time $\tau_\mathrm{QSL}$, which expresses the generalized energy-time uncertainty relation for arbitrary quantum unitary processes, in the generic form,
\begin{equation}
\label{q37}
\tau_\mathrm{QSL}=\ma{\frac{\hbar\,\mathcal{L}\left(\rho_\tau,\rho_0\right)}{E_\tau},\frac{\hbar\,\mathcal{L}\left(\rho_\tau,\rho_0\right)}{\Delta E_\tau}}\,.
\end{equation}
The above expression generalizes in an unified manner previous results derived in Refs.~\cite{ana90,vai91,uff93,bro03,man45,fle73,bha83,mar98,gio03,gio04,bur69,jon10,uhl92,pfe93,pfe95,lev09}. Similar formulas have found applications e.g. in the study of  lower bounds on the escape time of a particle out of a potential well modeling a quantum dot, or of the total time before which a $\mathrm{He}^+$ ion moving
in a uniform magnetic field loses its electron \cite{pfe95}. We first remark, as one might expect, that the quantum speed limit time is reduced when the Bures length between initial and final states is smaller, indicating that the closer the two states, the less time is needed to go from one state to the other. More importantly, the quantum speed limit time is inversely proportional to the energy of the system (mean energy or energy spread). This should not surprise as  energy is the generator of quantum time evolution. Whereas  $\tau_\mathrm{QSL}$ is determined by the initial energy in the time-independent case, it is set by the time-averaged mean energy or energy variance in the case of driven quantum systems. The quantum speed limit time, therefore, strongly depends on the driving protocol: the more energy is pumped into a system and the more it is distributed among different states, the faster  the system can evolve. For instance, if the quantum evolution time is controlled by the mean energy and the latter is increased exponentially in time, as in the creation of a squeezed state  in a modulated harmonic trap \cite{gal08}, the quantum speed limit time can be reduced exponentially. An extreme example is given by a quantum system initially in its pure, non-degenerate ground state. While the system cannot leave the initial (stationary) state in the undriven case as $\tau_\mathrm{QSL}$ is infinite, the quantum system does evolve with a finite dynamical speed once driven by an external parameter. On the other hand, we note that the quantum speed limit time can also be increased when the energy of the system is decreased by the driving \cite{def08}. The importance of the precise knowledge of the quantum speed limit time can be highlighted with a recent numerical study performed by Caneva {\it et al.} \cite{can09}. They showed that the Krotov algorithm of quantum optimal control theory fails to converge when the duration of the process is chosen below the quantum speed limit time. The latter thus represents the fundamental quantum limit of  the efficiency of the algorithm.
Finally, it is worth mentioning that  the existence of a quantum speed limit time is a purely quantum phenomenon, as it depends explicitly on the constant $\hbar$. In the classical limit, $\hbar \rightarrow 0$, the quantum speed limit vanishes, showing that classical systems can, in principle, evolve arbitrarily fast. However, as discussed by e.g. Landau and Lifshitz \cite{lan80}, if a classical system is driven so rapidly that the driving frequency becomes comparable to the typical transition frequency between quantum states,  quantum effects can no longer be neglected and have hence to be taken into account.

To conclude, we have derived extensions of the energy-time uncertainty relation for driven quantum systems that are valid for arbitrary initial and final states, as well as for arbitrary unitary driving. By determining upper bounds to the dynamical velocity, defined as the time derivative of the Bures length, we have established explicit expressions for the quantum speed limit time, that is, the minimal time a quantum system needs to evolve between two distinct states.  We have obtained generalizations of both the Mandelstam-Tamm  and Margolus-Levitin inequalities. Our result provides the fundamental limit on the evolution rate of  closed quantum systems, in particular quantum computation and  information processing devices.

\ack
The proof of Eq.~(\ref{q35}) is based on a suggestion made by an anonymous Referee. This work was supported by  the Emmy Noether Program of the DFG (contract No LU1382/1-1) and the cluster of excellence Nanosystems Initiative Munich (NIM).  SD also acknowledges financial support by a fellowship within the postdoc-program of the German Academic Exchange Service (DAAD, contract No D/11/40955).

\section*{Appendix}
A generalized Margolus-Levitin bound for time-dependent Hamiltonian can be derived following Ref.~\cite {jon10}  by considering the dynamical velocity  $\td_t\, {\cal L}$ itself. 
The time evolution of the quantum system obeys the time-dependent Schr\"odinger equation,
\begin{equation}
\label{q12}
i\hbar\,  \td_t\, \ket{{\psi_t}}=H_t\,\ket{\psi_t}\,.
\end{equation}
Taking the time derivative of the fidelity (\ref{q03}) for pure states, $
\mathcal{L}\left(\psi_0,\psi_t\right)=\arccos{\left(\left|\braket{\psi_0}{\psi_t} \right| \right)}$,
we find for the absolute value of the dynamical velocity,
\begin{eqnarray}
\label{q13}
\left|\td_t \mathcal{L}\right|=\frac{1}{\sqrt{1-\left|\braket{\psi_0}{\psi_t} \right|^2}}\,\left|\td_t\left|\braket{\psi_0}{\psi_t} \right|\right|=\frac{1}{\si{\mathcal{L}}}\,\left|\td_t\left|\braket{\psi_0}{\psi_t} \right|\right|\,.
\end{eqnarray}
Note that  $\td_t \mathcal{L}$ is positive at short times, since $\mathcal{L}$ is initially zero ($0\leq \mathcal{L}\leq \pi/2$). However, the derivative may change sign later on, and special care should be taken when deriving inequalities (this fact seems not have been noticed by Jones and Kok \cite{jon10}). In order to simplify Eq.~(\ref{q13}), we will prove that
\begin{equation}
\label{q14}
\left|\td_t\left|\braket{\psi_0}{\psi_t} \right|\right|\leq \left|\td_t\braket{\psi_0}{\psi_t} \right|\,.
\end{equation}
To this end, we expand the derivative on the left-hand side of Eq.~(\ref{q14}) as,
\begin{equation}
\label{q15}
\td_t\left|\braket{\psi_0}{\psi_t} \right|=\td_t\sqrt{\braket{\psi_0}{\psi_t}\braket{\psi_t}{\psi_0}}\,.
\end{equation}
Equation (\ref{q15}) can be evaluated  with the help of the time-dependent Schr\"odinger equation (\ref{q12}) to yield,
\begin{eqnarray}
\label{q16}
\left|\td_t\left|\braket{\psi_0}{\psi_t} \right|\right|&=&\left|\frac{\braket{\psi_0}{\psi_t}\bra{\psi_t}H_t\ket{\psi_0}- \bra{\psi_0}H_t\ket{\psi_t}\braket{\psi_t}{\psi_0}}{2i\hbar\,\left|\braket{\psi_t}{\psi_0} \right|}\right|\nonumber\\
&=&\left|\frac{\mathrm{Im}\left(\bra{\psi_t}H_t\ket{\psi_0}\braket{\psi_0}{\psi_t}\right)}{\hbar\,\left|\braket{\psi_t}{\psi_0} \right|}\right|\\
&\leq&\frac{\left|\bra{\psi_t}H_t\ket{\psi_0}\braket{\psi_0}{\psi_t} \right|}{\hbar\,\left|\braket{\psi_t}{\psi_0} \right|}\nonumber\,.
\end{eqnarray}
On the other hand,  $\left|\td_t\braket{\psi_0}{\psi_t} \right|$ can be  bounded from below by noting that,
\begin{equation}
\fl
\label{q17}
\left|\td_t\braket{\psi_0}{\psi_t} \right|=\frac{1}{\hbar}\left|\bra{\psi_0}H_t\ket{\psi_t} \right|=\frac{\left|\bra{\psi_t}H_t\ket{\psi_0} \right|\left|\braket{\psi_0}{\psi_t} \right|}{\hbar\,\left|\braket{\psi_t}{\psi_0} \right|}=\frac{\left|\bra{\psi_t}H_t\ket{\psi_0} \braket{\psi_0}{\psi_t} \right|}{\hbar\,\left|\braket{\psi_t}{\psi_0} \right|}\,.
\end{equation}
Comparing Eqs.~(\ref{q16}) and (\ref{q17})  leads to the desired estimation in Eq.~(\ref{q14}). 

\paragraph{Time-independent Hamiltonians.}

It has lately been recognized that the geometric derivation of the Margolus-Levitin bound put forward in Ref.~\cite{jon10} does not hold true in general \cite{zwi12}. For this reason, we first discuss the time-independent case. For constant Hamiltonians, the time-dependent solution of Eq.~(\ref{q12}) is given by,
\begin{equation}
\label{q18}
\ket{\psi_t}=\exp{\left(-itH/\hbar\right)}\,\ket{\psi_0}=\sum\limits_n \exp{\left(-itE_n/\hbar\right)}\, \alpha_n \ket{n} \ ,
\end{equation}
where $\alpha_n=\braket{n}{\psi_0}$. Note that the energy-eigenstates $\ket{n}$ are time-independent since $H$ is constant. Therefore we have,
\begin{eqnarray}
\fl
\left|d_t \braket{\psi_0}{\psi_t}\right|&=&\left|d_t \sum\limits_n \left|\alpha_n\right|^2\,\exp{\left(-\frac{it}{\hbar}\,E_n\right)}\right|\nonumber\\
\label{q19}&=&\left|\sum\limits_n\left|\alpha_n\right|^2\,\exp{\left(-\frac{it}{\hbar}\,E_n\right)}\left(-\frac{i}{\hbar}\,E_n\right)\right|\\
\label{q20}&\leq&\sum\limits_n\left|\left|\alpha_n\right|^2\,\exp{\left(-\frac{it}{\hbar}\,E_n\right)}\left(-\frac{i}{\hbar}\,E_n\right)\right|\\
\label{q21}&=&\frac{1}{\hbar}\,\sum\limits_n \left|\alpha_n\right|^2 E_n=\frac{1}{\hbar}\,\bra{\psi_0} H \ket{\psi_0}=\frac{1}{\hbar}\,\bra{\psi_t} H\ket{\psi_t} \ .
\end{eqnarray}
Combining Eqs.~(\ref{q13}), (\ref{q14}) and (\ref{q19})-(\ref{q21}), we find,
\begin{equation}
\label{q22}
\si{\mathcal{L}}\,\left|\td_t\mathcal{L}\right|\leq \left|\td_t\braket{\psi_0}{\psi_t} \right|=\frac{1}{\hbar}\left|\bra{\psi_t} H\ket{\psi_t}\right|\,.
\end{equation}
A final integration, with $|\int\td x\, f(x)|\leq\int\td x\,|f(x)|$ and $\si{x}\geq0$ for all $x\in[0,\pi/2]$, results in, 
\begin{equation}
\label{q23}
\left|\int_0^{{\cal L}\left(\psi_0,\psi_\tau\right)}\td{\cal L}\, \si{\mathcal{L}}\right|\leq\frac{1}{\hbar}\int_0^\tau \td t\, \left|\bra{\psi_t} H\ket{\psi_t}\right|\,.
\end{equation}
The average energy is positive since the ground state energy is  zero. We thus obtain,
\begin{equation}
\label{q24}
\tau\geq \frac{\hbar }{E_\tau}\,\left|\co{\mathcal{L}\left(\psi_0,\psi_\tau\right)}-1\right|\geq\frac{4\hbar}{\pi^2\,E_\tau}\,\mathcal{L}^2\left(\psi_0,\psi_\tau\right)\,,
\end{equation}
where we have used the trigonometric inequality, $\left|\co{x}-1\right|\geq(4/\pi^2)\,x^2$ for all $x\in[0,\pi/2]$. Here $E_\tau=(1/\tau)\, \int_0^\tau\td t\, \left|\bra{\psi_t} H \ket{\psi_t}\right|$ is the time averaged mean energy. Equation (\ref{q24}) is the Margolus-Levitin bound for quantum systems evolving under time-independent Hamiltonians, valid for arbitrary angles between initial and final states; it corrects previously published versions  \cite{jon10,zwi12}.

\paragraph{Time-dependent Hamiltonians.} 

For time-dependent Hamiltonians, the situation is more complicated. By combining Eqs.~(\ref{q13}), (\ref{q14}) and (\ref{q17}), we obtain,
\begin{equation}
\label{q25}
\si{\mathcal{L}}\,\left|\td_t\mathcal{L}\right|\leq \left|\td_t\braket{\psi_0}{\psi_t} \right|=\frac{1}{\hbar}\left|\bra{\psi_0} H_t\ket{\psi_t}\right|\,.
\end{equation}
In general, the instantaneous eigenbasis of $H_t$ is  time-dependent. As a result, the estimations in Eqs.~(\ref{q19})-(\ref{q21}) do not necessarily hold true. Under certain conditions, however, Eq.~(\ref{q24}) remains valid. Concretely,  the inequality,
\begin{equation}
\label{q26}
|\bra{\psi_0}H_t\ket{\psi_t}|\leq |\bra{\psi_t}H_t\ket{\psi_t}| \ ,
\end{equation} 
holds true if  $[H_t,H_0]=0\,,\hspace{1em} \forall\hspace{1em} t\in[0,\tau]$. This condition can be written as $\left|\braket{\psi_0}{n_t}\braket{n_t}{\psi_t}\right|\leq\left|\braket{\psi_t}{n_t}\braket{n_t}{\psi_t}\right|$, where $\{\ket{n_t}\}$ is the instantaneous energy eigenbasis with $H_t\ket{n_t}=E^t_n\ket{n_t}$. With  the latter,
 \begin{eqnarray}
\label{q27}
|\bra{\psi_0}H_t\ket{\psi_t}|&=&\left|\sum_n\,E_n^t\,\braket{\psi_0}{n_t}\braket{n_t}{\psi_t}\right|\\
\label{q28}&\leq&\sum_n\,E_n^t\,\left|\braket{\psi_0}{n_t}\braket{n_t}{\psi_t}\right|\\
\label{q29}&\leq&\sum_n\,E_n^t\,\left|\braket{\psi_t}{n_t}\braket{n_t}{\psi_t}\right|= |\bra{\psi_t}H_t\ket{\psi_t}|\ .
\end{eqnarray}
In more general cases, where Eq.~(\ref{q26}) is not valid, we can generalize Eq.~(\ref{q23}) to,\begin{equation}
\label{q30}
\left|\int_0^{{\cal L}\left(\psi_0,\psi_\tau\right)}\td{\cal L}\, \si{\mathcal{L}}\right|\leq\frac{1}{\hbar}\int_0^\tau \td t\, \left|\bra{\psi_0} H_t\ket{\psi_t}\right|\ ,
\end{equation}
by using  $|\int\td x\, f(x)|\leq\int\td x\,|f(x)|$ and $\si{x}\geq0$ for all $x\in[0,\pi/2]$. After time integration, we obtain,
\begin{equation}
\label{q31}
\tau\geq \frac{\hbar }{\tilde{E_\tau}}\,\left|\co{\mathcal{L}\left(\psi_0,\psi_\tau\right)}-1\right|\geq\frac{4\hbar}{\pi^2\,\tilde{E_\tau}}\,\mathcal{L}^2\left(\psi_0,\psi_\tau\right)\,,
\end{equation}
where we have again used the trigonometric inequality, $\left|\co{x}-1\right|\geq(4/\pi^2)\,x^2$ for all $x\in[0,\pi/2]$.  Here $\tilde{E_\tau}=(1/\tau)\, \int_0^\tau\td t\, \left|\bra{\psi_0} H_t U_t \ket{\psi_0}\right|$ is the time average of the nondiagonal matrix elements of the Hamiltonian, $|\bra{\psi_0} H_t\ket{\psi_t}|$   ($U_t$ is the time evolution operator).  Expression (\ref{q31}) is  a generalized Margolus-Levitin uncertainty relation valid  for arbitrary angles between initial and final pure states and arbitrary external driving.

The above derivation can be extended to  arbitrary mixed states by interpreting pure states as purifications of mixed states in an enlarged Hilbert space \cite{joz94}. Specifically, by choosing the purification that maximizes the fidelity between  two density operators, the Margolus-Levitin inequality   also  applies to the purifications \cite{jon10}. Hence, we generally have,
\begin{equation}
\label{q32}
\tau\geq\frac{4\hbar}{\pi^2\,\tilde{E_\tau}}\, \mathcal{L}^2\left(\rho_0,\rho_\tau\right) \ ,
\end{equation}
with the mean quantity $\tilde{E_\tau}=(1/\tau)\, \int_0^\tau\td t\, \spur{\rho_0 H_t U^\dagger_t}$. Equation (\ref{q32}) generalizes  the Margolus-Levitin-type bound introduced by Giovannetti, Lloyd and Maccone for time-independent systems \cite{gio03} (note that the numerical  prefactor $4/\pi^2$ in Eq.~(\ref{q32}) is smaller than the prefactor $2/\pi$ obtained numerically in Ref.~\cite{gio03}). In contrast to Eq.~(\ref{q36}), the quantum speed limit time (\ref{q32}) depends on the square of the Bures angle.


\section*{References}
\begin{thebibliography}{99}
\bibitem{aul09} G. Auletta, M. Fortunato and G. Parisi 2009 \textit{Quantum Mechanics}, (Cambridge, Cambridge)
\bibitem{hei27} W. Heisenberg 1927 \textit{Z. Phys.} \textbf{43} 172
\bibitem{boh28} N. Bohr 1928 \textit{Nature (Suppl.)} \textbf{121} 580
\bibitem{rob29} H. P. Robertson 1929 \PR \textbf{34} 163
\bibitem{bus07} P. Busch in: J. G. Mulga, R. Sala Mayato, and I. L. Egusquiza (Eds.) 2007 \textit{Time in Quantum Mechanics I} (Springer, Berlin) 
\bibitem{aha61} Y. Aharonov and D. Bohm 1961 \PR \textbf{122} 1649
\bibitem{ana90} J. Anandan and Y. Aharonov 1990 \PRL \textbf{65} 1697
\bibitem{vai91} L. Vaidman 1991 \textit{Am. J. Phys.} {\bf 60} 182
\bibitem{uff93} J. Uffink 1993 \textit{Am. J. Phys.} {\bf 61} 935
\bibitem{bro03} D. C. Brody 2003 \textit{J. Phys. A} \textbf{36} 5587
\bibitem{man45} L. Mandelstam and I. Tamm 1945 \textit{J. Phys. (USSR)} \textbf{9} 249
\bibitem{fle73} G. N. Fleming 1973 \textit{Nuovo Cimento A} \textbf{16} 232
\bibitem{bha83} K. Bhattacharyya 1983 \textit{J. Phys. A} \textbf{16} 2993
\bibitem{mar98} N. Margolus and L. B. Levitin 1998 \textit{Physica D} \textbf{120} 188
\bibitem{gio03} V. Giovannetti, S. Lloyd and L. Maccone 2003 \PRA \textbf{67} 052109
\bibitem{gio04} V. Giovannetti, S. Lloyd and L. Maccone 2004 \textit{J. Opt. B} \textbf{6} S807
\bibitem{bur69} D. J. C. Bures 1969 \textit{Trans. Amer. Math. Soc.} \textbf{135} 199
\bibitem{jon10} P. Jones and P. Kok 2010 \PRA \textbf{82} 022107
\bibitem{uhl92} A. Uhlmann 1992 \textit{Phys. Lett. A} \textbf{161}  329
\bibitem{pfe93} P. Pfeifer 1993 \PRL \textbf{70} 3365
\bibitem{pfe95} P. Pfeifer and J. Fr\"ohlich 1995 \RMP \textbf{67}, 759
\bibitem{lev09} L. B. Levitin and Y. Toffoli 2009 \PRL \textbf{103}, 160502
\bibitem{zwi10} M. Zwierz, C. A. P\'{e}rez-Delgado, and P. Kok 2010 \PRL \textbf{105} 180402
\bibitem{gio11} V. Giovannetti, S. lloyd, and L. Maccone 2011 \textit{Nature photonics} \textbf{5} 222
\bibitem{bek81} J. D. Bekenstein 1981 \PRL \textbf{46}, 623
\bibitem{llo00} S. Lloyd 2000 \textit{Nature (London)} \textbf{406}, 1047�1054
\bibitem{def10} S. Deffner and E. Lutz 2010 \PRL \textbf{105} 170402
\bibitem{can09} T. Caneva, M. Murphy, T. Calarco, R. Fazio, S. Montangero, V. Giovannetti, and G.E. Santoro 2009 \PRL \textbf{103} 240501
\bibitem{mur10} M. Murphy, S. Montangero, V. Giovannetti, and T. Calarco 2010 \PRA \textbf{82}, 022318
\bibitem{ben06} I. Bengtsson and K. \.{Z}yczkowski 2006 \textit{Geometry of Quantum States} (Cambridge, Cambridge)
\bibitem{woo81} W. Wootters 1981 \PRD \textbf{23} 357
\bibitem{cov06} T. M. Cover and J. A. Thomas 2006 \textit{Elements of Information Theory} (Wiley, New York)
\bibitem{kak48} S. Kakutani 1948 \textit{Ann. Math.} \textbf{49} 214
\bibitem{uhl76} A. Uhlmann 1976 \textit{Rep. Math. Phys.} \textbf{9} 273
\bibitem{joz94} R. Jozsa 1994 \textit{J. Mod. Opt.} \textbf{41} 2315
\bibitem{nie00} M. A. Nielsen and I. L. Chuang 2000 \textit{Quantum Computation and Quantum Information} (Cambridge, Cambridge)

\bibitem{bra94} S. Braunstein and C. Caves 1994 \PRL \textbf{72} 3439
\bibitem{hub92} M. H\"ubner 1992 Phys. Lett. A \textbf{163} 239
\bibitem{bra96} S. L. Braunstein, C. M. Caves, and G.J. Milburn 1996 \textit{Ann. Phys. (NY)} \textbf{247} 135 
\bibitem{gal08} F. Galve and E. Lutz 2009 \PRA \textbf{79} 055804
\bibitem{def08} S. Deffner and E. Lutz 2008 \PRE \textbf{77} 021128
\bibitem{lan80} L. D. Landau and E. M. Lifshitz 1980 \textit{Statistical Mechanics} Sect. 110 (Pergamon, Oxford) 
\bibitem{zwi12} M. Zwierz, 2012 \textit{Phys. Rev. A} \textbf{86}, 016101.
\end{thebibliography}
\end{document}